\documentclass[runningheads,]{llncs}
\usepackage[T1]{fontenc}
\usepackage{graphicx}
\usepackage{color}
\usepackage{xcolor}
\usepackage{xurl} 
\usepackage{hyperref}
\usepackage{booktabs}

\newcommand{\point}[1]{\par\smallskip\noindent\textbf{#1.}}
\begin{document}
\title{Pricing Factors and TFMs for Scalability-Focused ZK-Rollups}
%
%
\author{
Stefanos Chaliasos\inst{1,2}
\and 
Nicolas Mohnblatt\inst{3}
\and
Assimakis Kattis
\and
Benjamin Livshits\inst{1}
}
\institute{
Imperial College London, UK
\and
zkSecurity
\and
Geometry Research
}

\authorrunning{S. Chaliasos et al.}
%
%
\maketitle              
\begin{abstract}
ZK-Rollups have emerged as a leading solution for blockchain scalability, leveraging succinct proofs primarily based on ZKP protocols. This paper explores the design of transaction fee mechanisms (TFMs) for ZK-Rollups, focusing on how key components like sequencing, data availability~(DA), and ZK proving interact to influence cost structures. We outline the properties that a suitable TFM should possess, such as incentive compatibility and net profitability. In addition, we propose alternatives for TFMs, discuss trade-offs, and highlight open questions that require further investigation in the context of ZK-Rollups.

\keywords{Zero-Knowledge Proofs  \and Rollups \and Blockchain Scalability \and Mechanism Design, \and Transaction Fee Mechanism.}
\end{abstract}
\section{Introduction}

The growing adoption of blockchains, exemplified by networks like Ethereum, has highlighted the persistent challenge of scalability. As demand for decentralized applications increases, legacy Layer-1 (L1) blockchains face significant limitations in transaction throughput, resulting in congestion~\cite{zhou2020solutions}. To address these issues, Layer-2 (L2) solutions, such as rollups~\cite{sguanci2021layer}, have emerged, offloading transaction execution to faster chains while still relying on L1 for security.

Among these scalability solutions, ZK-Rollups~\cite{WhiteHat_roll_up_token} have become one of the most promising solutions. ZK-Rollups leverage ZKPs to produce validity proofs for L2 state transitions, which are then submitted for verification on L1. Despite their complexity, ZK-Rollups are considered the `holy grail' of L2 scalability due to their ability to provide fast finality and efficient data compression. However, while ZK-Rollups effectively address scalability, they introduce challenges in designing TFMs due to the intricate interplay between their various components.

In this paper we outline the design space for efficient and secure TFMs for ZK-Rollups that optimize operational costs while maintaining security and profitability. ZK-Rollups must balance several components, 
while existing literature provides insights into components individually~\cite{chaliasos2024analyzing}, a comprehensive analysis of how they interact within a TFM framework is still lacking.



\point{Contributions} 
\begin{itemize} 
    \item We outline the key components of ZK-Rollups, analyze their associated costs, and discuss how these factors influence transaction pricing (Section~\ref{sec:components}). 
    \item We define the desired properties of ZK-Rollups from the perspectives of both an L2 blockchain operator(s) and a transaction fee mechanism, examining the tradeoffs designers must consider (Section~\ref{sec:properties}). 
    \item We propose several alternative options for ZK-Rollup TFMs (Section~\ref{sec:mechanisms}).
    Finally, we offer initial insights and identify open questions for future research required in designing TFMs for ZK-Rollups that can be optimized based on specific utility functions set by system designers (Section~\ref{sec:future}). 
\end{itemize}

\section{Main Components of ZK-Rollups}
\label{sec:components}

This section outlines the key components of ZK-Rollups, each critical to the rollup’s cost structure. Some components may be integrated or optimized differently across implementations. Table~\ref{tab:components} summarizes their roles.

\begin{table}[t]
\centering
\resizebox{\columnwidth}{!}{ 
\begin{tabular}{lll}
\toprule
\textbf{Component} & \textbf{Description} & \textbf{Properties Affected} \\ 
\midrule
Settlement & Verifies proofs on the L1. & CE, F1, OC \\ 
Data Availability & Ensures transaction data or state diffs are accessible for L1 verification. & OC \\
Sequencer & Orders and executes L2 transactions. & CS, F2, EV \\ 
Batch Sealing & Aggregates L2 blocks into a batch for L1 submission. & F1, EV \\ 
L2 Consensus & Provides soft-finality in L2. & F2, CS \\ 
ZK Prover & Generates validity proofs for a batch of transactions. & OC \\ 
\bottomrule\\
\end{tabular}
}
\caption{Components and properties of ZK-Rollups. Correct Execution: CE, CS: Censorship Resistance, EV: Economic Value, OC: Operational Costs, F1: Finality (L1), F2: Finality (L2).}
\vspace{-0.5cm}
\label{tab:components}
\end{table}

\point{C1 - Settlement~---~L1 Verification}
The L1 chain serves as the settlement layer for ZK-Rollups, verifying the proofs submitted by the rollup and ensuring the integrity of L2 transactions. Chains like Ethereum validate these proofs, making L1 settlement a significant cost contributor. Settlement costs are fixed per batch, meaning the total cost is divided among the L2 transactions in the batch. Hence, smaller batches result to higher cost per L2 transaction.

\point{C2 - Data Availability (DA)}
DA is essential for ZK-Rollups to ensure that all required transaction data becomes accessible for verification in the L1. DA was a major bottleneck for rollups that leveraged Ethereum as an L1 since all data needed to be published as calldata, which is extremely costly in periods of high demand. The introduction of blobs to Ethereum alleviates this bottleneck by allowing cheaper DA for rollups. Although DA is currently not a limiting cost, that could change in case of high demand for storing data on-chain. 

\point{C3 - Sequencer}
The sequencer is one of the most critical components in ZK-Rollups, responsible for ordering and executing transactions in L2s~\cite{motepalli2023SoKDS}. There are three main sequencing models:
\emph{(i)} \textit{Self-sovereign sequencing}: The rollup operates its own sequencer,
\emph{(ii)} \textit{Shared sequencing}: Multiple rollups share a common sequencer, and
\emph{(iii)} \textit{`Based' rollups}: Sequencing tasks are partially delegated to L1.
The role of the sequencer is vital not only because it can extract MEV if allowed, but also since it determines which transactions are included in each L2 block and, consequently, in each batch.
Beyond its traditional roles, the sequencer in ZK-Rollups can also act as an optimizer for reducing the total cost of a rollup batch.
For example, in Polygon's zkEVM, the prover's workload remains constant, whether it is proving one or multiple instances of a complex operation such as keccak hashing. Therefore, to maximize efficiency, the sequencer should group transactions that approach the prover's processing limit for the specific expensive-to-prove operation into the same batch. Similarly, in rollups submitting state diffs in DA, the sequencer can play a pivotal role in the DA cost. By batching transactions that interact with the same contracts, sequencers can reduce DA costs by minimizing state diffs, which lowers the overall bytes required to be submitted in the L1.
However, these optimizations may conflict with each other and with other goals, like extracting priority fees or MEV, leading to tradeoffs that must be carefully managed and even dynamically adjusted based on exogenous changes (e.g., an increase in DA cost).

\point{C4 - Batch Sealing}
Batch sealing is the process of aggregating multiple L2 blocks into a single batch to be submitted on L1. The criteria for batch sealing can vary, including time intervals, the number of transactions, or prover limitations. Like sequencing, batch sealing can significantly influence both DA and prover costs. Moreover, it must balance the tradeoff between cost amortization and finality speed, especially during periods of low usage -- waiting for more transactions to reduce costs or submitting batches more frequently for faster finality. In many rollups, batch sealing may be integrated within the sequencer.

\point{C5 - L2 Consensus}
Different L2s employ various consensus mechanisms, often tailored for low latency and simplicity compared to L1. In L2 systems, consensus typically provides soft-finality (i.e., finality on the L2) for transactions, with full finality achieved on L1 when the validity proof of a batch is verified, and secured by the consensus of the L1.
Most ZK-Rollups today use proof-of-authority (PoA) in centralized and trusted sequencers. When trusted nodes are available, proof-of-efficiency (PoE) can be employed, which focuses on ensuring high efficiency by using trusted entities to achieve consensus on the rollup with minimal overhead~\cite{proof_of_efficiency}.
For decentralized rollups or shared-sequenced rollups, HotShot has been proposed~\cite{bearer2024espresso}. 
It is particularly suitable for rollups because of its ability to handle network fluctuations and provide quick finality with lower overhead than traditional L1 consensus systems. 
In summary, ZK-Rollup consensus is simpler than L1, focusing on fast, low-cost, and low-latency soft finality, which preserves scalability with minimal impact on costs.

\point{C6 - ZK Prover}
The ZK prover generates zero-knowledge proofs that verify the correctness of the rollup’s transactions. Key steps are: \emph{(i)} \textit{proving}, which refers to the initial proof generation; \emph{(ii)} \textit{aggregation/recursion}, where multiple proofs are combined into a single, smaller proof; and \emph{(iii)} \textit{compression}, which involves techniques for transitioning from prover-friendly (e.g. STARK, GKR) to verification-friendly (e.g. Groth16, Fflonk) proof systems to reduce proof size, impacting verification time and cost, albeit with additional proving costs.
The efficiency of the ZK prover directly influences transaction fees in ZK-Rollups, as fees must cover proving costs. Since most rollups are currently centralized, an open challenge remains in incentivizing and fairly compensating provers. Furthermore, in some rollups that use Ethereum's gas metering mechanism, there is a risk of DoS attacks on the prover, as proving costs may significantly differ from the execution costs captured by Ethereum's gas model.

\section{Properties}
\label{sec:properties}

ZK-Rollups' key properties, such as security, censorship resistance, finality, and costs, are shaped by the interaction of components like the sequencer, prover, and L1 settlement. In this section, we introduce these properties, explore how each component influences them, and discuss the tradeoffs involved in balancing them. Table~\ref{tab:components} summarizes which properties are affected by each component.

\point{P1 - Correct Execution}
The security of ZK-Rollups is primarily derived from the validity proofs that ensure the correct execution of L2 batches, as well as from the L1, which verifies these proofs. 
For an in-depth analysis of ZK-Rollup security, we refer the reader to~\cite{chaliasos2024towards}. 
An L2 can opt for an alternative settlement layer, such as a verification service,~\footnote{\url{https://alignedlayer.com/}} to reduce costs at the expense of different security guarantees. However, these solutions are outside the scope of this paper.

\point{P2 - Censorship Resistance \& Decentralization}
A rollup inherits censorship resistance from the L1 by enabling direct L2 transactions via L1,\footnote{Note that this feature it's not currently implemented by all rollups.} though these are significantly more expensive. Decentralizing the L2 sequencer and consensus can provide censorship resistance at the L2 level, but it is still constrained (i.e., upper bounded) by L1 censorship, as L1 can refuse to process certain batches. Decentralizing the prover is also important, as centralized provers could censor transactions, though doing so would require censoring an entire batch. While decentralization improves fault tolerance, centralized designs can also achieve some fault tolerance, so this aspect is beyond the scope of our work.

\point{P3 - Finality}
ZK-Rollups have a dual finality: \emph{(i)} \textit{L2 finality (soft-finality)}, achieved when a transaction is included in an L2 block, typically in under 2 seconds, and \emph{(ii)} \textit{L1 finality (hard-finality)}, achieved when the batch containing the L2 transaction is committed to L1, its proof verified, and the L1 block is finalized. L2 finality is mainly influenced by the Sequencer and Consensus, while L1 finality depends on Settlement, Sequencer, Batch Sealing, and the ZK Prover. There is a tradeoff between faster finality and higher settlement cost per transaction when using smaller, faster-to-prove batches.

\point{P4 - Operational Costs \& Economic Value}
Some L2 components, like the ZK Prover, incur significant operational costs due to high energy consumption and hardware requirements, while L1 DA and settlement costs fluctuate based on L1 prices and demand. These costs are also influenced by external factors such as global energy prices, cloud services, and demand for on-chain data storage. In contrast, components like the Sequencer have minimal operational costs but high economic value due to their control over transaction inclusion, ordering, and potential MEV extraction, making them a low-cost, high-value asset in the TFM. Similarly, Batch Sealing has low costs but high value, as it impacts expensive components like the ZK Prover and DA. Finally, we assume that Consensus in rollups incurs neither significant operational costs nor high economic value.

\point{P5 - TFM Desired Properties} An ideal TFM for a ZK-Rollup should exhibit several desirable properties:

\par\smallskip\noindent\emph{(1)} \textbf{Net profitability}: The rollup should cover its operational costs and generate income through transaction fees and MEV, except in certain edge cases. 
\par\smallskip\noindent\emph{(2)} \textbf{DoS attack protection}: The fee structure must protect the ZK Prover from potential DoS attacks, as proving costs may differ from execution costs as captured by Ethereum’s gas metering mechanism. 
\par\smallskip\noindent\emph{(3)} \textbf{Incentive compatibility}: Ensuring both User Incentive Compatibility (UIC) and Miner Incentive Compatibility (MIC)~\cite{roughgarden2021tfm,chung2021FoundationsOT}. 
\par\smallskip\noindent\emph{(4)} \textbf{Off-chain agreement (OCA) proofness}: Preventing off-chain collusion between users, sequencers, or provers. 
\par\smallskip\noindent\emph{(5)} \textbf{Fees predictability (optional)}: Due to the fluctuating nature of L1 costs (DA and Settlement), predicting fees from the time of L2 transaction to batch settlement on L1 can be challenging. One solution is to overcharge users and refund the difference, which leads to a poor user experience. An alternative approach is deploying futures markets on L1 to hedge against price fluctuations~\cite{tsabary2022ledgerhedger}.



\section{Potential Alternatives for TFM for ZK-Rollups}
\label{sec:mechanisms}
In this section, we explore three orthogonal dimensions for designing TFMs for ZK-Rollups. First, we examine the suitability of multidimensional fee markets for ZK-Rollups, followed by an analysis of the tradeoffs between centralized and decentralized TFMs. Lastly, we discuss the pros and cons of using subsidies to cover operational costs. Each approach requires further research and empirical evaluation, particularly when considering combinations of these mechanisms.

\point{Multidimensional Fee Markets}
Multidimensional fee markets~\cite{diamandis2023designing,crapis2023OptimalDF} present a promising approach for ZK-Rollups, as they manage multiple resources with varying costs, such as sequencing, DA, and proving. Each commodity is priced independently based on demand and consumption, offering a more efficient and accurate reflection of actual costs. For ZK-Rollups, this allows fine-tuned pricing that accounts for the complexity of proving costs and unpredictable L1 expenses, enabling the decoupling of component costs to optimize outcomes like minimizing prover costs or improving finality speed. Further, multidimensional fee markets are particularly promising for achieving net profitability, as they allow precise alignment of costs with resource usage.


\point{Centralized vs. Decentralized TFM}
Most ZK-Rollups currently use centralized TFMs, offering advantages like near-zero consensus costs and flexible, real-time adjustments to optimize efficiency and reduce operational expenses. The centralized operator, has perfect knowledge, and can maximize efficiency and improve the user experience. However, this approach has drawbacks: censorship resistance relies solely on L1, which is costly, and centralized sequencers may extract MEV, risking reputational damage if seen as unfair. Further, these systems depend heavily on the operator’s trustworthiness. In contrast, decentralized TFMs involve varying degrees of permission per market/component, and designers must carefully consider the tradeoffs between decentralization degrees.


\point{Subsidized TFM}
An alternative approach to TFMs for ZK-Rollups could involve subsidizing some operational costs. In this model, certain costs, such as proving or DA, are partially covered by the protocol itself. The goal would be to create a more competitive rollup that attracts higher usage, which in turn leads to increased fee revenue and greater MEV extraction potential. By offering lower fees, the rollup could gain market share, and the system could profit not only through direct fees but also through the appreciation of its native token, which in turn can be used to further compensate the rollup operators.
This approach raises several challenges and ethical concerns, as it could lead to distortion in fee structures and user incentives. The long-term sustainability of subsidized systems is also questionable, as it could create dependence on inflation or market conditions. Further, this strategy may incentivize short-term gains over long-term network security and stability, which requires further investigation.


\section{Future Work \& Open Questions}
\label{sec:future}
In this section, we outline critical open questions and areas for further research in the design of TFMs for ZK-Rollups.

\point{Metering Mechanisms and DoS Risk}
A key area for future research is designing a metering mechanism that accurately captures all relevant costs within ZK-Rollups. Current metering mechanisms, particularly those for EVM-compatible VMs, inherit Ethereum's metering approach. This could pose a risk of DoS attacks, as these mechanisms may fail to account for differences between execution costs (captured by gas fees) and proving costs (which are unique to ZK-Rollups). Future research should focus on evaluating whether this discrepancy can be exploited and how to improve metering mechanisms to reflect the actual resource consumption in ZK-Rollups.

\point{Comprehensive Mechanism Design}
Future work should focus on developing comprehensive TFMs that integrate all components of ZK-Rollups. The goal should be to optimize different properties, such as cost efficiency, finality, and security, while balancing the various trade-offs between these components. This could involve dynamic pricing models, batch optimization, and incentive structures that align the goals of different actors in the system, including provers and sequencers.

\point{Incentive Compatibility}
Ensuring incentive compatibility between different actors, such as provers and sequencers, remains a challenge. Future research should explore models that align the incentives of all actors involved with the overall performance and security of the ZK-Rollup. For example, provers should be incentivized to minimize proving costs, while sequencers might be driven by MEV extraction or transaction inclusion fees. Mechanisms must be designed to ensure that these incentives do not conflict and do not undermine system performance or security.

\point{Cost Management Strategies}
Another open question is how to effectively manage costs in periods of low usage while optimizing transaction finality. Potential strategies include:
\begin{itemize}
    \item Dynamic Pricing: Implementing a dynamic pricing model where transaction costs rise during periods of low usage to encourage more activity and better cost amortization.
    \item Batch Optimization: Adjusting batch sizes according to network activity. For instance, smaller batches can be used during periods of low activity for fast finality, while larger batches are used during peak times to reduce per-transaction costs.
    \item Flexible Finality: Allowing users to choose between fast finality with higher costs or delayed finality with lower costs. This would give users more control over their transaction costs and priorities.
    \item Blob-Space Sharing and Calldata Optimization: Research could explore the potential of blob-space sharing and selectively using calldata instead of blobs to optimize costs based on current network conditions and blob prices.
\end{itemize}

\point{Quantitative Feedback and Implementation}
We also aim to implement and test a mechanism in practice to evaluate its effectiveness. Collecting quantitative feedback from real-world implementations or simulations would provide valuable insights into how different TFMs perform in practice. These empirical data could guide future optimization efforts and help refine mechanisms based on actual performance metrics, such as throughput, finality times, and cost savings.

\point{Understanding Trade-offs}
Exploring the tradeoffs between various design choices in ZK-Rollups is crucial for their long-term viability. Questions to consider include:

\begin{itemize}
    \item Can a ZK-Rollup remain sustainable without subsidies, or is some level of subsidy required for long-term success?
    \item How can the rollup's revenue, especially from MEV, be maximized without negatively affecting users? Should rollups focus on maximizing direct revenue through fees, or should they prioritize TVL (Total Value Locked) growth, even with minimal fees, to attract more transactions and fees?
    \item How should the user experience be optimized, particularly in terms of fee structures? Should users be overcharged to guarantee safety, with potential refunds later, or should alternative approaches be explored?
    \item Does decentrilization have a negative effect on TFMs?
\end{itemize}

\point{Non-Rollup Solutions}
Another area to explore is the implications of non-rollup solutions, such as using external DA services or a separate verification layer. These alternatives may offer cost improvements, but could affect the security guarantees provided by L1. Understanding the trade-offs between cost efficiency and security in these non-rollup solutions could offer new avenues for optimizing scalability designs that use validity proofs beyond ZK-Rollups.

\section{Related Work}
\label{app:related}
\point{Transaction Fee Mechanism Design}
Traditional Transaction Fee Mechanisms for Layer-1 blockchains, such as Ethereum, initially relied on first-price auctions, which proved inefficient and susceptible to off-chain collusion. The introduction of EIP-1559 sought to resolve these issues by combining a base fee and burn mechanism and enhancing miner and user incentives. Roughgarden~\cite{roughgarden2021tfm} offers a comprehensive analysis of various mechanisms, introduces the Off-Chain Agreement (OCA) collusion-resistance notion (preventing collusion between miners and users off-chain), and explores the nuances of the TFM space, particularly EIP-1559. Chung and Shi~\cite{chung2021FoundationsOT} contribute by defining Side-Contract Proofness (SCP) (resistance to collusions between miners and users) and demonstrating the impossibility of achieving both Dominant Strategy Incentive Compatibility (DSIC) (where a user’s best strategy is to bid truthfully, regardless of others' actions) and the SCP mechanisms.
Recent research has proposed multidimensional fee markets as a solution. Diamandis et al.\cite{diamandis2023designing} and Crapis et al.\cite{crapis2023OptimalDF} introduce dynamic pricing models, where resources like gas and bandwidth are priced separately, improving allocation efficiency and mitigating risks like denial-of-service (DoS) attacks. Similarly, proposed mechanisms for ZK-Rollups could be based on multi-dimensional fee markets.

The most relevant work to ours is Wang et al.~\cite{wang2024mechanism}'s Proo$\phi$ mechanism, which offers an innovative auction model for ZK-Rollup prover markets, focusing on provers' compensation and collusion resistance. While their work makes significant strides, we extend this work by integrating all critical components of ZK-Rollups, emphasizing the importance of considering costs and resource demands holistically. Our paper particularly highlights the role of sequencing and optimized batching to minimize proving costs, along with a detailed exploration of the various trade-offs designers must account for in ZK-Rollup systems.

\point{Benchmarking and Analyzing ZK Systems}
Orthogonal to TFM, papers analyzing and benchmarking ZKPs are highly relevant to our work, as understanding how provers perform in practice is key to fine-tuning TFMs for ZK-Rollups. The most notable contribution in this space is by Chaliasos et al.\cite{chaliasos2024analyzing}, who analyzed ZK-Rollups both qualitatively and quantitatively. They categorized ZK-Rollup costs into fixed and marginal, then benchmarked different systems, providing valuable insights into prover performance. Their findings showed how various ZK-Rollups handle different workloads and resource demands, demonstrating the significant impact of sequencing transactions on ZK-Rollups' efficiency. Further studies have focused on benchmarking ZKP frameworks\cite{ernstberger2024zkbench} and hash functions for EVM-compatible blockchains~\cite{Guo2024BenchmarkingZH}. These works complement our research by highlighting the importance of optimizing the proving utility in ZK-Rollups, ensuring that TFM designs account for practical performance constraints and resource demands.

\section{Conclusion}
This paper explored the key components of ZK-Rollups and their impact on transaction fee mechanisms. We analyze how components such as data availability, sequencing, and ZK proving affect cost structures and outline important trade-offs in TFM design, such as finality speed versus cost efficiency. We discussed  orthogonal dimensions of TFMs, including multidimensional fee markets and subsidized TFMs, and highlighted open questions for further research.

\begin{credits}
\subsubsection{\ackname} We thank the Latest in DeFi Research (TLDR) program,
funded by the Uniswap Foundation, for supporting this work.

\end{credits}
%
%
\bibliographystyle{splncs04}
\bibliography{main}

\end{document}